\documentclass[twocolumn,prb,superscriptaddress,noshowpacs,epsf]{revtex4}
\usepackage{graphicx}

\begin{document}
\title{Quantum information processing with superconducting qubits in a microwave field}
%
%
\author{J. Q. You}
\affiliation{Frontier Research System, The Institute of Physical
and Chemical Research (RIKEN), Wako-shi 351-0198, Japan}
\affiliation{Center for Theoretical Physics, Physics Department,
Center for the Study of Complex Systems,
The University of Michigan, Ann Arbor, MI 48109-1120, USA}
\affiliation{National Laboratory for Superlattices and Microstructures,
Institute of Semiconductors, Chinese Academy of Sciences, Beijing 100083,
China}
\author{Franco Nori}
\altaffiliation[Corresponding author~(Email:~nori@umich.edu).]{}
\affiliation{Frontier Research System, The Institute of Physical and
Chemical Research (RIKEN), Wako-shi 351-0198, Japan}
\affiliation{Center for Theoretical Physics, Physics Department, Center
for the Study of Complex Systems, The University of Michigan, Ann Arbor,
MI 48109-1120, USA}
\altaffiliation[Permanent address.]{}

\begin{abstract}
We investigate the quantum dynamics of a Cooper-pair box with a
superconducting loop in the presence of a nonclassical microwave field.
We demonstrate the existence of Rabi oscillations for both
single- and multi-photon processes
and, moreover, we propose a new quantum computing scheme
(including one-bit and conditional two-bit gates) based on Josephson qubits
coupled through microwaves.
\end{abstract}
\maketitle

\section{Introduction}
\subsection{Background}

Quantum computing deals with the processing of information according to
the laws of quantum mechanics. Within the last few years, it has attracted
considerable attention because quantum computers are expected to be
capable of performing certain tasks which no classical computers can
do in practical time scales.
Early proposals for quantum computers were mainly based
on quantum optical systems, such as those utilizing laser-cooled
trapped ions,\cite{CIR,MON} photon or atoms in quantum elctrodynamical
(QED) cavities,\cite{SLE,TUR} and nuclear magnetic resonance.\cite{NIE}
These systems are well isolated from their environment and satisfy the
low-decoherence criterion for implementing quantum computing.
Moreover, due to quantum error correction algorithms,\cite{NIE} now
decoherence~\cite{HAN} is not regarded as an insurmountable barrier to
quantum computing.
Because scalability of quantum computer architectures to many
qubits is of central importance for realizing quantum computers of
practical use, considerable efforts have recently been devoted to
solid state qubits.
Proposed solid state architectures include those using electron
spins in quantum dots,\cite{LOS,IMA,XHU} electrons on
Helium,\cite{PLA} and Josephson-junction (JJ) charge (see, e.g.,
Refs.~\onlinecite{SHN,AV,FAL} and \onlinecite{MAK})
and JJ flux (see, e.g., Refs.~\onlinecite{MOO} and \onlinecite{MAK}) devices.
These qubit systems have the
advantage of relatively long coherent times and are expected to be
scalable to large-scale networks using modern microfabrication techniques.

The Josephson charge qubit is achieved in a Cooper-pair box,\cite{SHN}
which is a small superconducting island weakly coupled to a bulk
superconductor, while the Josephson flux qubit is based on two
different flux states in a small
superconducting-quantum-interference-device (SQUID) loop.\cite{MOO,MAK}
Cooper-pair tunneling and energy-level splitting associated with
the superpositions of charge states were experimentally demonstrated
in a Cooper-pair box,\cite{NAK1,BOU} and recently the eigenenergies and
the related properties of the superpositions of different flux states were
observed in SQUID loops by spectroscopic measurements.\cite{VAN} In
particular, Nakamura {\it et al.}~\cite{NAK2} demonstrated the quantum
coherent oscillations of a Josephson charge qubit prepared in a
superposition of two charge states. In addition, Vion {\it et al.}~\cite{VION}
extended coherent oscillations to the charge-flux regime and
Chiorescu {\it et al.}~\cite{CHIO} studied the quantum dynamics of the flux qubit.
Moreover, two charge qubits were capacitively coupled by
Pashkin {\it et al.}~\cite{PASH} and coherent oscillations
were also observed in this coupled-qubit system. Furthermore, other superconducting
devices (e.g., Refs.~\onlinecite{YU} and \onlinecite{MART}) have also exhibited
coherent oscillations.
In addition, several other types of studies (see, for instance,
Refs.~\onlinecite{others} and \onlinecite{YTN}) have been made on
superconducting qubits.

\subsection{This work}

In this paper, we show that the coupled system of a Cooper-pair box
and a cavity photon mode undergoes Rabi oscillations and propose a
new quantum computing scheme based on Josephson charge qubits.\cite{flux}
The microwave-controlled approach proposed in our paper has the significant
advantage that {\it any} two qubits ({\it not} necessarily neighbors)
can be effectively coupled through photons in the cavity.
In addition to the advantages of a superconducting device exhibiting
quantum coherent effects in a macroscopic scale as well as the
controllable feature of the Josephson charge qubit by {\it both} gate
voltage {\it and} external flux, the motivation for this scheme is
fourfold:

(i) the experimental measurements~\cite{NAK1} showed that the energy
difference between the two eigenstates in a Cooper-pair box
lies in the microwave region and the eigenstates can be
effectively interacted by the microwave field;

(ii) a single photon can be readily prepared in a {\it high-Q} QED cavity
using the Rabi precession in the microwave domain.\cite{MAI} Moreover,
using a QED cavity, Ref.~\onlinecite{BRA} produced a reliable source
of photon number states on demand. In addition, the cavity in Ref.~\onlinecite{BRA}
was tuned to $\sim 21$~GHz, which is close to the $20$~GHz microwave
frequency used in a very recent experiment~\cite{NAK3} on the Josephson
charge qubit. Furthermore,
the $Q$ value of the cavity is $4\times 10^{10}$ (giving a very large
photon lifetime of $0.3$~sec);

(iii) our quantum computer proposal
should be scalable to $10^6$ to $10^8$ charge qubits in a microwave cavity,
since the dimension of a Cooper-pair box is $\sim 10\mu$m to $ 1 \mu$m;

(iv) the QED cavity has the advantage that {\it any} two qubits
({\it not} necessarily neighbors) can be effectively coupled through
photons in the cavity.

Also, we study multi-photon processes in the Josephson charge
qubit since, in contrast to the usual Jaynes-Cummings
model (see, {\it e.g.}, Chap.~10 in Ref.~\onlinecite{WAL}),
the Hamiltonian includes higher-order interactions
between the two-level system and the nonclassical microwave field.
As shown by the very recent experiment on Rabi oscillations in a
Cooper-pair box,\cite{NAK3} these higher-order interactions may be
important in the Josephson charge-qubit system.

Note that the driving microwave field is typically generated using
an {\it electrical\/} voltage acting on the charge qubit via a
gate capacitor. Here, the microwave field is applied as a {\it
magnetic\/} flux piercing the SQUID loop of the qubit in order to
perform the unitary transformations needed for quantum computing.

The dynamics of a Josephson charge qubit coupled to a quantum
resonator was studied in Ref.~\onlinecite{BUI}. In contrast to our
study here, the model in Ref.~\onlinecite{BUI} involves: (a)~ only
one qubit, (b) only the Rabi oscillation with a single excitation
quantum of the resonator (as opposed to one or more photons), and
(c) {\it no\/} quantum computing scheme.

\section{Charge Qubit in a Cavity}
\subsection{Cooper-pair box with a SQUID loop}

We study the Cooper-pair box with a SQUID loop.\cite{SHN,MAK,NAK2}.
In this structure, the
superconducting island with Cooper-pair charge $Q=2ne$ is coupled to a
segment of a superconducting ring via two Josephson junctions
(each with capacitance $C_J$ and Josephson coupling energy $E_{J0}$).
Also, a voltage
$V_g$ is coupled to the superconducting island through a gate capacitor
$C_g$; the gate voltage $V_g$ is externally controlled and used to induce
offset charges on the island. A schematic illustration of  this
single-qubit structure is given in Fig.~1(a). The Hamiltonian of the system is
\begin{equation}
H=4E_c\left(n-{C_gV_g\over 2e}\right)^2-E_J(\Phi)\cos\varphi,
\end{equation}
where
\begin{equation}
E_c={e^2\over 2(C_g+2C_J)}
\end{equation}
is the single-particle charging energy of the
island and
\begin{equation}
E_J(\Phi)=2E_{J0}\cos\left({\pi\Phi\over\Phi_0}\right)
\end{equation}
is the effective
Josephson coupling. The number $n$ of the extra Cooper pairs on the island
and average phase drop
\[
\varphi={1\over 2}(\varphi_1+\varphi_2)
\]
are canonically
conjugate variables. The gauge-invariant phase drops $\varphi_1$ and
$\varphi_2$ across the junctions are related to the total flux
$\Phi$ through the SQUID loop by the constraint
\begin{equation}
\varphi_2-\varphi_1=2\pi{\Phi\over\Phi_0},
\end{equation}
where $\Phi_0=h/2e$ is the flux
quantum.

\begin{figure}
\includegraphics[width=3.0in,height=2.4in,bbllx=120,bblly=365,
bburx=533,bbury=680]{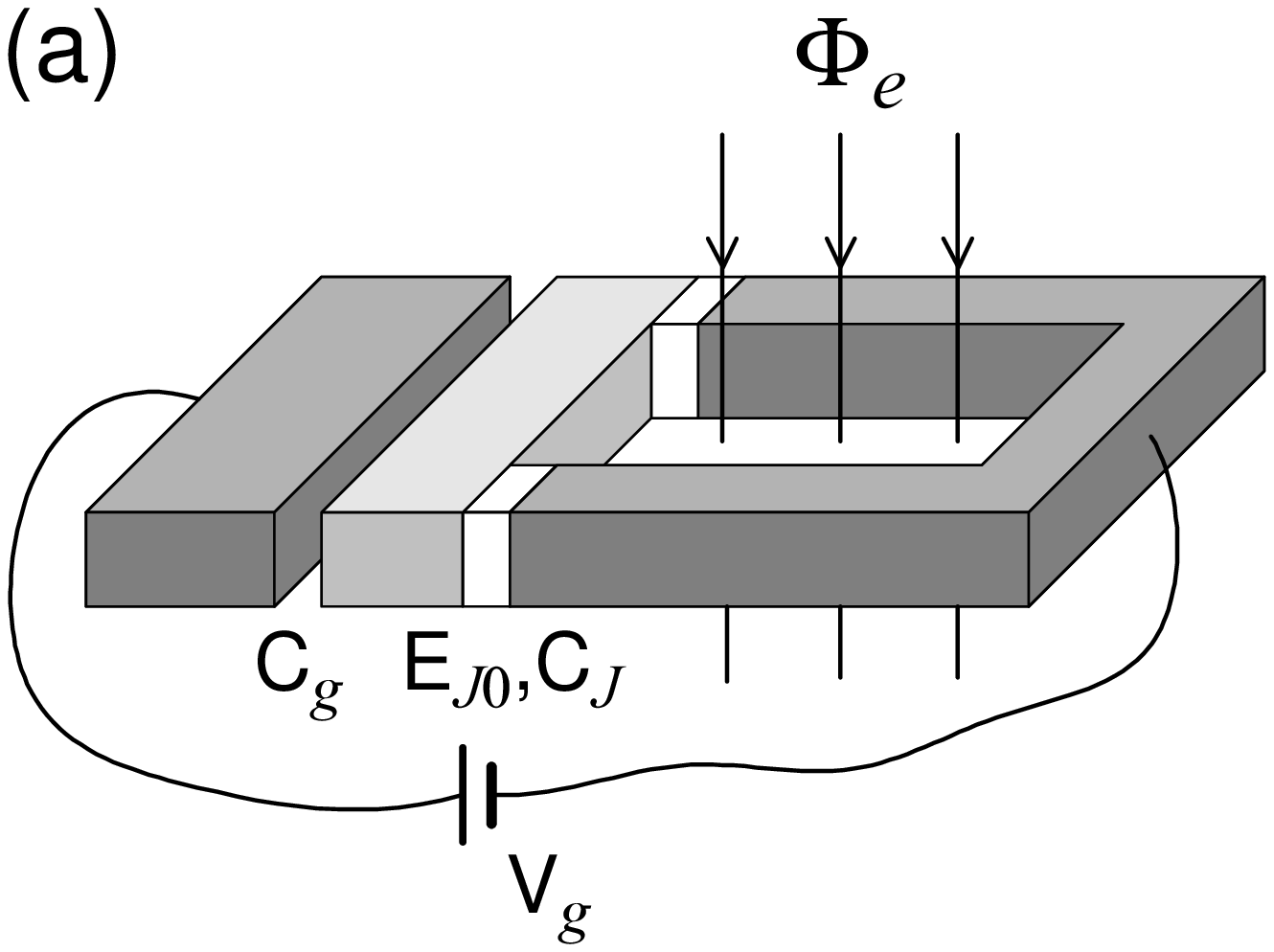}
\includegraphics[width=3.0in,height=2.4in,bbllx=120,bblly=365,
bburx=533,bbury=680]{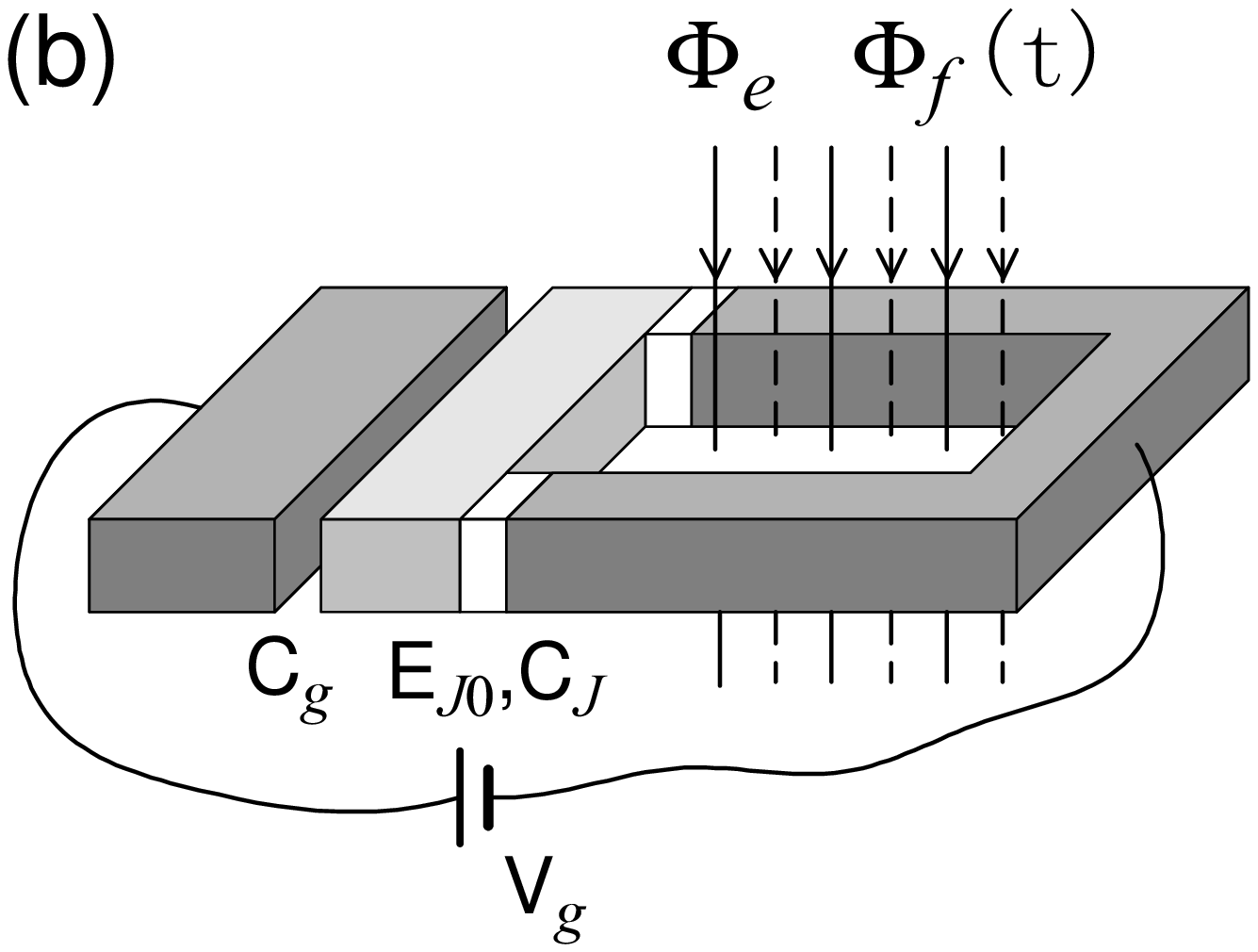}
%
%
\caption{Cooper-pair box with a SQUID loop, where the charge box
is coupled to a segment of a superconducting ring via two
identical Josephson junctions, shown in white above, and a voltage
$V_g$ is applied to the charge box through a gate capacitor $C_g$
(on the left side of the above diagram). (a) A static magnetic
flux $\Phi_e$, as denoted by the solid lines with arrows, pierces
the SQUID loop to control the effective Josephson coupling energy.
(b) In addition to $\Phi_e$, a microwave field $\Phi_f(t)$,
schematically shown above by the dashed lines with arrows, is also
applied through the SQUID loop.} \label{fig1}
\end{figure}

This structure is characterized by two energy scales, i.e., the
charging energy $E_c$ and the coupling energy $E_{J0}$ of the Josephson
junction.
In the charging regime $E_c\gg E_{J0}$ and at low temperatures $k_BT\ll E_c$,
the charge states $|n\rangle$ and $|n+1\rangle$ become dominant as the
{\it controllable} gate voltage is adjusted to $V_g\sim (2n+1)e/C_g$.
Here, the superconducting gap is assumed to be larger than $E_c$,
so that quasiparticle tunneling is greatly reduced in the system.

Here we ignore self-inductance effects on the single-qubit
structure.\cite{YOU}
Now $\Phi$ reduces to the classical variable $\Phi_e$, where $\Phi_e$ is the
flux generated by the applied {\it static} magnetic field.
In the spin-${1\over 2}$ representation with charge states
$|\!\uparrow\rangle=|n\rangle$ and $|\!\downarrow\rangle=|n+1\rangle$, the
reduced two-state Hamiltonian is given by~\cite{SHN,MAK}
\begin{equation}
H=\varepsilon(V_g)\,\sigma_z-{1\over 2}E_J(\Phi_e)\sigma_x,
\end{equation}
where
\begin{equation}
\varepsilon(V_g)=2E_c\left[{C_gV_g\over e}-(2n+1)\right].
\end{equation}
This single-qubit amiltonian has two eigenvalues
\begin{equation}
E_{\pm}=\pm{1\over 2} E,
\end{equation}
with
\begin{equation}
E=\left[4\varepsilon^2(V_g)+E_J^2(\Phi_e)\right]^{1/2},
\end{equation}
and eigenstates
\begin{eqnarray}
|e\rangle\!&\!=\!&\!\cos\xi\:|\!\uparrow\rangle
-\sin\xi\:|\!\downarrow\rangle, \nonumber\\
|g\rangle\!&\!=\!&\!\sin\xi\:|\!\uparrow\rangle
+\cos\xi\:|\!\downarrow\rangle,
\end{eqnarray}
with
\begin{equation}
\xi={1\over 2}\tan^{-1}\left({E_J\over 2\varepsilon}\right).
\end{equation}
Using these eigenstates as new basis, the Hamiltonian takes the diagonal form
\begin{equation}
H={1\over 2}E\rho_z,
\end{equation}
where
\begin{equation}
\rho_z=|e\rangle\langle e|-|g\rangle\langle g|.
\end{equation}
Here we employ $\{|e\rangle,|g\rangle\}$ to represent the qubit.

\subsection{Interaction of the charge qubit with a microwave field}

When a {\it nonclassical} microwave field is applied, the total flux
$\Phi$ is a quantum variable
\begin{equation}
\Phi=\Phi_e+\Phi_f(t),
\end{equation}
where $\Phi_f$ is the microwave-field-induced flux through the SQUID
loop [see Fig.~1(b)]. Here we assume that a single-qubit structure is
embedded in a QED microwave
cavity with only a single photon mode $\lambda$. Generally, the vector
potential of the nonclassical microwave field is written as
\begin{eqnarray}
{\bf A}({\bf r})\!&\!=\!&\!{\bf u}_{\lambda}({\bf r})a
+{\bf u}_{\lambda}^*({\bf r})a^{\dag} \nonumber\\
&\!=\!&\!|{\bf u}_{\lambda}({\bf r})|(e^{-i\theta}a+e^{i\theta}a^{\dag})\hat{\bf A},
\end{eqnarray}
where $a^{\dag} (a)$ is the creation (annihilation) operator of the cavity
mode. Thus, the flux $\Phi_f$ is given by
\begin{equation}
\Phi_f=|\Phi_{\lambda}|(e^{-i\theta}a+e^{i\theta}a^{\dag}),
\end{equation}
with
\begin{equation}
\Phi_{\lambda}=\oint{\bf u}_{\lambda}\cdot d{\bf l},
\end{equation}
where the contour integration is over the SQUID loop.
Here, $\theta$ is the phase of the mode function
${\bf u}_{\lambda}({\bf r})$ and its value depends on the chosen
microwave field (see, {\it e.g.}, Chap.~2 in Ref.~\onlinecite{WAL}).
For instance, if a planar cavity is used and the SQUID loop of the charge
qubit is perpendicular to the cavity mirrors, one has $\theta=0$.

We shift the gate voltage $V_g$ (and/or vary $\Phi_e$) to bring the
single-qubit system into
resonance with $k$ photons:
\begin{equation}
E\approx k\hbar\omega_{\lambda},\;\;\; k=1,2,3,\dots.
\end{equation}
Expanding the functions $\cos(\pi\Phi_f/\Phi_0)$
and $\sin(\pi\Phi_f/\Phi_0)$ into series of operators and employing the
standard rotating wave approximation, we derive the total Hamiltonian of
the system in this situation (with the photon Hamiltonian included)
\begin{eqnarray}
&&H={1\over 2}E\rho_z+\hbar\omega_{\lambda}\left(a^{\dag}a+{1\over
2}\right)+H_{Ik}, \label{hami}\\
&&H_{Ik}=\rho_zf(a^{\dag}a)+\left[e^{-ik\theta}
|e\rangle\langle g|a^kg^{(k)}(a^{\dag}a)+{\rm H.c.}\right]. \nonumber
\end{eqnarray}
Here
\begin{equation}
f(a^{\dag}a)=-E_{J0}\sin(2\xi)\cos\left({\pi\Phi_e\over\Phi_0}\right)F(a^{\dag}a),
\end{equation}
with
\begin{eqnarray}
F(a^{\dag}a)\!&\!=\!&\!{1\over
2!}\phi^2(2a^{\dag}a+1)
-{3\over 4!}\phi^4\left[2(a^{\dag}a)^2+2a^{\dag}a+1\right] \nonumber\\
&&\!+{5\over 6!}\phi^6
\left[4(a^{\dag}a)^3+6(a^{\dag}a)^2+8a^{\dag}a+3\right] \nonumber\\
&&\!-\dots,
\end{eqnarray}
where $\phi=\pi|\Phi_{\lambda}|/\Phi_0$, and
\begin{eqnarray}
&&g^{(2m-1)}(a^{\dag}a)=E_{J0}\cos(2\xi)\sin\left({\pi\Phi_e\over\Phi_0}\right)
G^{(2m-1)}(a^{\dag}a), \nonumber\\
&&g^{(2m)}(a^{\dag}a)=E_{J0}\cos(2\xi)\cos\left({\pi\Phi_e\over\Phi_0}\right)
G^{(2m)}(a^{\dag}a),
\end{eqnarray}
with $m=1,2,3,\dots$, and
\begin{eqnarray}
&&G^{(1)}(a^{\dag}a)=\phi-{1\over 2!}\phi^3a^{\dag}a
+{1\over 4!}\phi^5\left[2(a^{\dag}a)^2+1\right]-\dots,\nonumber\\
&&G^{(2)}(a^{\dag}a)={1\over 2!}\phi^2-{2\over 4!}\phi^4
\left(2a^{\dag}a-1\right) \nonumber\\
&&~~~~~~~~~~~~~~~+{15\over 6!}\phi^6
\left[(a^{\dag}a)^2-a^{\dag}a+1\right]-\dots,\nonumber\\
&&G^{(3)}(a^{\dag}a)=-{1\over 3!}\phi^3
+{5\over 5!}\phi^5\left(a^{\dag}a-1\right)-\dots,\nonumber\\
&&G^{(4)}(a^{\dag}a)=-{1\over 4!}\phi^4
+{3\over 6!}\phi^6\left(2a^{\dag}a-3\right)-\dots,\nonumber\\
&&\dots\dots\dots,
\end{eqnarray}
where $g^{(k)}(a^{\dag}a)$ is the $k$-photon-mediated coupling between the
charge qubit and the microwave field. This Hamiltonian~(\ref{hami})
is a generalization of the Jaynes-Cummings model to
a solid state system. Here multi-photon processes~\cite{MEE} are
involved for $k>1$, in contrast with the usual
Jaynes-Cummings model for an atomic two-level system interacting with a
single photon mode, where only one photon is exchanged between the
two-level system and the external field.\cite{WAL}

\section{Rabi Oscillations in Multi-Photon Process}

The eigenvalues of the total Hamiltonian~(\ref{hami}) are
\begin{eqnarray}
{\cal E}_{\pm}(l,k)\!&\!=\!&\!\hbar\omega_{\lambda}\left[l+{1\over 2}(k+1)\right]
+{1\over 2}\left[f(l)-f(l+k)\right] \nonumber\\
&&\!\pm {\hbar\over 2}\sqrt{\delta_{l,k}^2+\Omega_{l,k}^2},
\end{eqnarray}
and the corresponding eigenstates, namely, the dressed states are given by
\begin{eqnarray}
|+,l\rangle\!&\!=\!&\!e^{-ik\theta}\cos\eta\:|e,l\rangle
+\sin\eta\:|g,l+k\rangle, \nonumber\\
|-,l\rangle\!&\!=\!&\!-\sin\eta\:|e,l\rangle+e^{ik\theta}\cos\eta\:|g,l+k\rangle,
\end{eqnarray}
where
\begin{equation}
\Omega_{l,k}=2g^{(k)}(l+k)\left[(l+1)(l+2)\cdots (l+k)\right]^{1/2}/\hbar
\end{equation}
is the Rabi frequency,
\begin{equation}
\delta_{l,k}=(E/\hbar-k\omega_{\lambda})+[f(l)+f(l+k)]/\hbar,
\end{equation}
and
\begin{equation}
\eta={1\over 2}\tan^{-1}\left({\Omega_{l,k}\over\delta_{l,k}}\right).
\end{equation}
Here, $k$ is the number of photons emitted or absorbed by the charge qubit
when the qubit transits between the excited state $|e\rangle$ and
the ground state $|g\rangle$, and $l$ is the number of photons in the cavity
when the qubit state is $|e\rangle$.

When the system is
initially at the state $|e,l\rangle$, after a period of time $t$, the
probabilities for the system to be at states $|g,l+k\rangle$ and
$|e,l\rangle$ are
\begin{equation}
|\langle g,l+k|\psi(t)\rangle|^2=
{\Omega_{l,k}^2\over\delta_{l,k}^2+\Omega_{l,k}^2}
\sin^2\left[{1\over 2}\left(\delta_{l,k}^2+\Omega_{l,k}^2\right)^{1/2}t\right],
\end{equation}
and
\begin{equation}
|\langle e,l|\psi(t)\rangle|^2=1-|\langle g,l+k|\psi(t)\rangle|^2.
\end{equation}
Thus, the probabilities are oscillating with
frequency
\begin{equation}
\Omega_{\rm Rabi}=\left(\delta_{l,k}^2+\Omega_{l,k}^2\right)^{1/2}.
\end{equation}
This is the
{\it Rabi oscillation with $k$ photons} involved in the state
transition; when $k=1$, it reduces to the usual single-photon Rabi
oscillation.

Very recently, Nakamura {\it et al.}~\cite{NAK3} investigated
the temporal behavior of a
Cooper-pair box driven by a strong microwave field and observed
the Rabi oscillations with {\it multi}-photon exchanges between
the two-level system and the microwave field. Different to the
case studied here, the microwave field was employed there to drive
the gate voltage to oscillate. Here, in order to implement quantum
computing, we consider the Cooper-pair box with a SQUID loop and
use the microwave field to change the flux through the loop.

\subsection{Analogies between Rabi oscillations and the AC Josephson effect}

Rabi oscillations have been observed a long time ago in atomic physics. It is a
relatively new development to observe Rabi oscillations in a condensed matter system.
Since the Josephson effect can be used for this purpose, it is instructive to
point out analogies and differences between Rabi oscillations and the Josephson effect.

(i)~Both Rabi oscillations and the AC Josephson effect involve interactions of the
photons with electrons (for Rabi oscillations) or a junction (for AC Josephson effect);
(ii)~the radiation must be
tuned creating two-level transitions; (iii)~the junction behaves like an
atom undergoing transitions between the quantum states of each side of the
junction as it absorbs and emits radiation.

However, the Rabi oscillation is a
strong-coupling effect~\cite{WAL} and produces long-lived coherent
superpositions.

\section{Quantum Computing}

Let us consider more than one single
charge qubit in the QED cavity, and the cavity initially prepared at the
zero-photon state $|0\rangle$. We first show the implementation of a
controlled-phase-shift operation.
Here a single photon process, $k=1$, is used to implement quantum computing.

(i) For all Josephson charge qubits, let
\[
\Phi_e={1\over 2}\Phi_0,
\]
then $\cos\left({\pi\Phi_e/\Phi_0}\right)=0$, which yields
\[
f(a^{\dag}a)=0.
\]
Furthermore, the gate voltage for a control qubit,
say $A$, is adjusted to have the qubit on resonance with the cavity mode
($E=\hbar\omega_{\lambda}$) for a period of time (where single photon is
involved in the state transition), while all other qubits are kept
off-resonant. The interaction Hamiltonian (in the interaction picture with
$H_0={1\over 2}E\rho_z$) is given by
\begin{equation}
H_{\rm int}=e^{-i\theta}|e\rangle_A\langle g|\,a\,g^{(1)}(a^{\dag}a)+{\rm
H.c.},
\end{equation}
and the evolution of qubit $A$ is described by
\begin{equation}
U_A(\theta,t)=\exp(-iH_{\rm int}t/\hbar).
\end{equation}
This unitary operation does not affect state
$|g\rangle_A|0\rangle$, but transforms $|g\rangle_A|1\rangle$ and
$|e\rangle_A|0\rangle$ as
\begin{eqnarray}
|g\rangle_A|1\rangle\!&\!\longrightarrow\!&\!\cos(\alpha t)|g\rangle_A|1\rangle
-ie^{-i\theta}\sin(\alpha t)|e\rangle_A|0\rangle, \nonumber\\
|e\rangle_A|0\rangle\!&\!\longrightarrow\!&\!\cos(\alpha t)|e\rangle_A|0\rangle
-ie^{i\theta}\sin(\alpha t)|g\rangle_A|1\rangle,
\end{eqnarray}
where $\alpha=g^{(1)}(1)/\hbar$. To obtain the controlled-phase-shift
gate, we need the unitary operation with $\theta=0$ and interaction time
$t_1=\pi/2\alpha$, which gives
\begin{eqnarray}
|g\rangle_A|1\rangle\!&\!\longrightarrow\!&\!-i|e\rangle_A|0\rangle,\nonumber\\
|e\rangle_A|0\rangle\!&\!\longrightarrow\!&\!-i|g\rangle_A|1\rangle.
\end{eqnarray}
This operation swaps the qubit state and the state of the QED cavity. A
similar swapping transformation was previously used for the quantum
computing with laser-cooled trapped ions.\cite{CIR}

(ii) While all qubits are kept off-resonant with the cavity mode and the
flux $\Phi_e$ is originally set to $\Phi_e={1\over 2}\Phi_0$
for each qubit, we change $\Phi_e$ to zero for only the target qubit, say $B$.
In this case, the evolution of the target qubit $B$ is described in the
interaction picture by
\begin{equation}
U_B(t)=\exp(-iH_{\rm int}t/\hbar),
\end{equation}
where the Hamiltonian is
\begin{equation}
H_{\rm int}=\left(|e\rangle_B\langle e|
-|g\rangle_B\langle g|\right)\,f(a^{\dag}a).
\end{equation}
This Hamiltonian can be used to produce {\it conditional} phase shifts
in terms of the photon state of the QED cavity.\cite{SLE}
Applying this unitary operation to qubit $B$ for a period of
time $t_2=\pi\hbar/2|f(1)-f(0)|$, we have~\cite{teleportation}
\begin{eqnarray}
&&|g\rangle_B|0\rangle\longrightarrow e^{i\beta}|g\rangle_B|0\rangle,\nonumber\\
&&|e\rangle_B|0\rangle\longrightarrow e^{-i\beta}|e\rangle_B|0\rangle,\nonumber\\
&&|g\rangle_B|1\rangle\longrightarrow ie^{i\beta}|g\rangle_B|1\rangle,\nonumber\\
&&|e\rangle_B|1\rangle\longrightarrow -ie^{-i\beta}|e\rangle_B|1\rangle,
\end{eqnarray}
where $\beta=\pi f(0)/2|f(1)-f(0)|$.

(iii) Qubit $A$ is again brought into resonance for
$t_3=\pi/2\alpha$ with $\theta=0$, as in step (i).
Afterwards, a controlled two-bit
gate is derived as a controlled-phase-shift gate combined with
two one-bit phase gates. In order to obtain
the controlled-phase-shift
gate $U_{AB}$, which transforms $|g\rangle_A|g\rangle_B$,
$|g\rangle_A|e\rangle_B$, $|e\rangle_A|g\rangle_B$, and
$|e\rangle_A|e\rangle_B$ as
\begin{equation}\left(
\begin{array}{c}
|g\rangle_A|g\rangle_B\\
|g\rangle_A|e\rangle_B\\
|e\rangle_A|g\rangle_B\\
|e\rangle_A|e\rangle_B
\end{array}\right)
\longrightarrow\left(
\begin{array}{cccc}
1 & 0 & 0 & 0 \\
0 & 1 & 0 & 0 \\
0 & 0 & 1 & 0 \\
0 & 0 & 0 & -1
\end{array}\right)
\left(
\begin{array}{c}
|g\rangle_A|g\rangle_B\\
|g\rangle_A|e\rangle_B\\
|e\rangle_A|g\rangle_B\\
|e\rangle_A|e\rangle_B
\end{array}\right),
\end{equation}
one needs to further apply successively the unitary operation given in
step (ii) to the control and target qubits with interaction times
$t_4=3\pi\hbar/4|f(0)|$ and $t_5=(2\pi-|\beta|)\hbar/|f(0)|$, respectively.

In analogy with atomic two-level systems,\cite{CIR,SLE} one can use an
appropriate {\it classical microwave field}$\:$~\cite{rotation} to produce
{\it one-bit rotations} for the Josephson
charge qubits. When the classical microwave field is on resonance with the
target qubit $B$, the interaction Hamiltonian becomes
\begin{equation}
H_{\rm int}\,=\,{\hbar\Omega\over
2}[e^{-i\nu}|e\rangle_B\langle g|+{\rm H.c.}],
\end{equation}
with
\begin{equation}
\hbar\Omega=2E_{J0}\cos(2\xi)\sin\left({\pi\Phi_e\over\Phi_0}\right)
\left({\pi|\Phi_f|\over\Phi_0}\right),
\end{equation}
where the value of the phase $\nu$ depends on the chosen microwave
field (see, {\it e.g.}, Chap.~2 in Ref.~\onlinecite{WAL}) and
$\Phi_f$ is the flux through the
SQUID loop produced by the classical microwave field. For the
interaction time $t_6=\pi/2\Omega$, the unitary operation
\begin{equation}
V_B(\nu,t_6)=\exp(-iH_{\rm int}t_6/\hbar)
\end{equation}
transforms $|g\rangle_B$
and $|e\rangle_B$ as
\begin{eqnarray}
&&|g\rangle_B\longrightarrow{1\over\sqrt 2}
\left(|g\rangle_B-ie^{i\nu}|e\rangle_B\right),\nonumber\\
&&|e\rangle_B\longrightarrow{1\over\sqrt 2}
\left(|e\rangle_B-ie^{-i\nu}|g\rangle_B\right).
\end{eqnarray}
In terms of this one-bit rotation, the controlled-phase-shift gate
$U_{AB}$ can be converted to the controlled-NOT gate,\cite{CIR}
\begin{equation}
C_{AB}=V_B\left(-{\pi\over 2},{\pi\over 2\Omega}\right)\,U_{AB}\,V_B
\left({\pi\over 2},{\pi\over 2\Omega}\right).
\end{equation}
A sequence of such gates supplemented by one-bit rotations can serve as
a universal element for quantum computing.\cite{LLO}

\section{Discussion and Conclusion}

For microwaves of wavelength $\lambda\sim 1$ cm, the volume of a planar cavity is
$\sim 1$cm$^3$. For SQUID loop dimension $\sim 10\mu$m to $1 \mu$m,
then $10^3$ to $10^4$ charge qubits
may be constructed along the cavity direction. Furthermore, for a 2D array of
qubits, $10^6$ to $10^8$ charge qubits could be placed within
the cavity.\cite{array} This number of qubits is large enough for a quantum
computer. For practical quantum information processing, one needs to improve
the experimental setup to have a QED cavity with a high enough Q value so as to
implement more quantum operations within the long photon lifetime of the cavity.
Alternatively, one can also increase the number of permitted quantum operations within
the given photon lifetime of the cavity by strengthening the coupling between
the charge qubit and the microwave field.
Because the typical interaction energy between
the charge qubit and the microwave field is propositional to $\Phi_{\lambda}$,
the qubit-photon coupling can be strenghthened by increasing the area enclosed by
the SQUID loop and the field intensity
(e.g. by putting a high-$\mu$  material inside the SQUID loop).

In the conditional gates discussed above, the two charge qubits
are coupled through photons in the QED cavity. Our approach is
scalable, but similar to the coupling scheme using an $LC$-oscillator
mode,\cite{SHN,MAK} only a pair of charge qubits at a time can be
coupled.
In order to implement parallel operations on different pairs of
qubits, one can make use of a multi-mode QED cavity or
more than one cavity, where {\it different
cavity modes couple different pairs of qubits simultaneously}.
Moreover, our approach
might have potential applications in quantum communications using
both the qubit-photon coupling (to convert quantum information
between charge qubits and photons) and the photons, acting as
flying qubits, to transfer quantum information between remotely
separated charge-qubit systems.

In conclusion, we have studied the dynamics of the Cooper-pair box with a
SQUID loop in the  presence of a nonclassical microwave field. Rabi
oscillations in the
multi-photon process are demonstrated, which involve multiple photons in
the transition between the two-level system  and the microwave
field. Also, we propose a scheme for quantum computing, which is
realized by Josephson charge qubits coupled through photons
in the QED cavity.

\begin{acknowledgments}
We thank X. Hu, B. Plourde, C. Monroe and C. Kurdak for useful comments, and
acknowledge support from ARDA, AFOSR, the US National Science
Foundation grant No.~EIA-0130383, and the National Natural Science
Foundation of China.
\end{acknowledgments}
%



\begin{references}
\bibitem{CIR} J.I.~Cirac and P.~Zoller, Phys.~Rev.~Lett.~{\bf 74}, 4091
(1995).
\bibitem{MON} C.~Monroe, D.M. Meekhof, B.E. King, W.M. Itano, and
D.J. Wineland, Phys.~Rev.~Lett.~{\bf 75}, 4714
(1995).
\bibitem{SLE} T.~Sleator and H.~Weinfurter, Phys.~Rev.~Lett.~{\bf 74}, 4087
(1995).
\bibitem{TUR} Q.A.~Turchette, C.J. Hood, W. Lange, H. Mabuchi, and H.J. Kimble,
Phys.~Rev.~Lett.~{\bf 75}, 4710
(1995).
\bibitem{NIE} M.A.~Nielsen and I.L.~Chuang, {\it Quantum Computation and
Quantum Information} (Cambridge University Press, Cambridge, 2000).
\bibitem{HAN} See, {\it e.g.}, M.~Thorwart and P.~H{\" a}nggi, Phys. Rev. A
{\bf 65}, 012309 (2001); J.~Shao and P.~H{\" a}nggi, Phys. Rev. Lett. {\bf 81},
5710 (1998).
\bibitem{LOS} D.~Loss and D.P.~DiVincenzo, Phys.~Rev.~A {\bf 57}, 120
(1998);
X.~Hu and S.~Das Sarma, Phys.~Rev.~A {\bf 61}, 062301 (2000), and
references therein.
\bibitem{IMA} A.~Imamo\={g}lu, D.D. Awschalom, G. Burkard, D.P. DiVincenzo, D. Loss,
M. Sherwin, and A. Small, Phys.~Rev.~Lett.~{\bf 83}, 4204
(1999); M.S. Sherwin, A. Imamo\={g}lu, and T. Montroy, Phys. Rev. A
{\bf 60}, 3508 (1999).
\bibitem{XHU} X.~Hu, R.~de Sousa, and S.~Das Sarma, Phys. ~Rev. ~Lett.
~{\bf 86}, 918
(2001).
\bibitem{PLA} P.M.~Platzman and M.I.~Dykman,
Science {\bf 284}, 1967 (1999).
%
%
\bibitem{SHN} A. Shnirman, G. Sch\"{o}n, and Z. Hermon,
Phys. Rev. Lett. {\bf 79}, 2371 (1997); Y. Makhlin, G. Sch\"{o}n,
and A. Shnirman, Nature (London) {\bf 398}, 305 (1999)
\bibitem{AV} D.V. Averin, Solid State Comm. {\bf 105}, 659 (1998)
\bibitem{FAL} G.~Falci, R. Fazio, G. M. Palma, J. Siewert, and V. Vedral,
Nature (London) {\bf 407}, 355 (2000);
%
A.~Blais and A.-M.S.~Tremblay, Phys. Rev. A {\bf 67}, 012308
(2003);
L. Faoro, J. Siewert, and R. Fazio, Phys. Rev. Lett. {\bf 90}, 028301 (2003).
%
\bibitem{MOO} J.E.~Mooij, T.P. Orlando, L. Levitov, L. Tian, C.H. van der Wal,
and S. Lloyd, Science {\bf 285}, 1036 (1999);
T.P.~Orlando, J.E. Mooij, L. Tian, C.H. van der Wal, L.S. Levitov, S. Lloyd,
and J.J. Mazo, Phys.~Rev.~B {\bf 60} 15398 (1999).
\bibitem{MAK} For a review on both Josephson charge and
flux qubits, see Y. Makhlin, G. Sch\"{o}n, and A. Shnirman,
Rev. Mod. Phys. {\bf 73}, 357 (2001).
\bibitem{NAK1} Y.~Nakamura, C.D.~Chen, and J.S.~Tsai, Phys. ~Rev. ~Lett.
~{\bf 79}, 2328 (1997).
\bibitem{BOU} V.~Bouchiat, D. Vion, P. Joyez, D. Esteve, and M.H. Devoret,
Phys.~Scripta {\bf T76}, 165 (1998).
\bibitem{VAN} C.H.~van der Wal, A.C.J. ter Haar, F.K. Wilhelm, R.N. Schouten,
C.J.P.M. Harmans, T.P. Orlando, S. Loyd, and J.E. Mooij, Science {\bf 290}, 773
(2000);
J.R.~Friedman, V. Patel, W. Chen, S.K. Tolpygo, and J.E. Lukens,
Nature (London) {\bf 406}, 43 (2000).
\bibitem{NAK2} Y.~Nakamura, Yu.~A.~Pashkin, and J.S.~Tsai, Nature (London)
{\bf 398}, 786 (1999).
\bibitem{VION} D. Vion, A. Aassime, A. Cottet, P. Joyez, H. Pothier, C. Urbina,
D. Esteve, and M.H. Devoret, Science {\bf 296}, 886 (2002).
\bibitem{CHIO} I. Chiorescu, Y. Nakamura, C.J.P.M. Harmans, and J.E. Mooij,
Science {\bf 299}, 1869 (2003).
\bibitem{PASH} Yu. A. Pashkin, T. Yamamoto, O. Astafiev, Y. Nakamura, D.V. Averin,
and J.S. Tsai, Nature (London) {\bf 421}, 823 (2003).
\bibitem{YU} Y. Yu, S. Han, X. Chu, S. Chu, and Z. Wang, Science {\bf 296}, 889 (2002).
\bibitem{MART} J.M. Martinis, S. Nam, J. Aumentado, and C. Urbina, Phys. Rev. Lett.
{\bf 89}, 117901 (2002).
%
\bibitem{others}
%
A.A.~Clerk, S.M.~Girvin, A.K.~Nguyen, and A.D.~Stone
Phys. Rev. Lett. {\bf 89}, 176804 (2002);
%
J. Siewert and R.~Fazio, {\it ibid.} {\bf 87}, 257905 (2001);
%
%
O.~Buisson, F.~Balestro, J.P.~Pekola, and F.W.J.~Hekking
{\it ibid.} {\bf 90}, 238304 (2003);
%
%
A.~Blais, A.~Maassen van den Brink, and A.M.~Zagoskin
{\it ibid.} {\bf 90}, 127901 (2003);
%
%
E. Bibow, P. Lafarge, and L.P. L{\'e}vy,
{\it ibid.} {\bf 88}, 017003 (2002);
%
E. Paladino, L. Faoro, G. Falci, and R. Fazio, {\it ibid.} {\bf 88}, 228304 (2002);
%
L.~Tian, S.~Lloyd, and T.P.~Orlando
Phys. Rev. B {\bf 65}, 144516 (2002);
%
%
G.~Blatter, V.B.~Geshkenbein, and L.B.~Ioffe
{\it ibid.} {\bf 63}, 174511 (2001);
%
%
%
%
J.M. Martinis, S. Nam, J. Aumentado, K.M. Lang, and C. Urbina
{\it ibid.} {\bf 67}, 094510 (2003);
%
%
P.R.~Johnson, F.W. Strauch, A.J. Dragt, R.C. Ramos, C.J.
Lobb, J.R. Anderson, and F.C. Wellstood
{\it ibid.} {\bf 67}, 020509 (2003);
%
%
%
%
%
Z.~Zhou, S.-I.~Chu, and S.~Han,
{\it ibid.} {\bf 66}, 054527 (2002);
%
%
S.~Oh, {\it ibid.} {\bf 65}, 144526 (2002);
%
E. Almaas and D. Stroud, {\it ibid} {\bf 65}, 134502 (2002);
%
%
V. Sch{\" o}llmann, P. Agren, D.B. Haviland, T.H. Hansson, and A.
Karlhede, {\it ibid.} {\bf 65}, 020505 (2002);
%
M.C.~Goorden and F.K.~Wilhelm, {\it ibid.}, in press;
%
%
T.L. Robertson, B.L.T. Plourde, A. Garcia-Martinez, P.A.
Reichardt, B. Chesca, R. Kleiner, Y. Makhlin, G. Sch{\"o}n, A.
Shnirman, F.K. Wilhelm, D.J. Van Harlingen, and J. Clarke, {\it
ibid.}, in press;
%
C.-P.~Yang, S.-I.~Chu, and S.~Han, Phys. Rev. A {\bf 67}, 042311
(2003);
%
D.V. Averin and C. Bruder, cond-mat/0304166;
%
M.J.~Storcz and F.K. Wilhelm, cond-mat/0306317.
%
%
\bibitem{YTN} A scalable superconducting qubit structure is proposed in
J.Q. You, J.S. Tsai, and F. Nori, Phys. Rev. Lett. {\bf 89},
197902 (2002). A longer version of it is available in
cond-mat/0306208.
%
Entanglement of states in a circuit with superconducting qubits
and improved readout is described in J.Q. You, J.S. Tsai, and F.
Nori, Phys. Rev. B {\bf 68}, 024510 (2003).
%
%
%
%
%
%
%
\bibitem{flux} While here we focus on charge qubits, similar ideas also apply
to {\it flux} qubits.
\bibitem{MAI} X.~Ma\^{i}tre, E. Hagley, G. Nogues, C. Wunderlich, P. Goy, M. Brune,
J.M. Raimond, and S. Haroche, Phys.~Rev.~Lett.~{\bf 79}, 769 (1997).
\bibitem{BRA} S.~Brattke, B.T.H.~Varcoe, and H.~Walther, Phys. Rev. Lett.
{\bf 86}, 3534 (2001).
\bibitem{NAK3} Y. Nakamura, Yu. A. Pashkin, and J.S. Tsai,
Phys. Rev. Lett. {\bf 87}, 246601 (2001).
\bibitem{WAL} See, {\it e.g.}, D.F.~Walls and G.J.~Milburn, {\it Quantum
Optics} (Springer, Berlin, 1994).
\bibitem{BUI} A.D. Armour, M.P. Blencowe, and K.C. Schwab, Phys. Rev. Lett.
{\bf 88}, 148301 (2002); O. Buisson and F.W.J.~Hekking, cond-mat/0008275,
in {\it Macroscopic Quantum Coherence and Quantum Computing} (Kluwer Academic,
Dordrecht), pp.~137-145;
F. Marquardt and C. Bruder, Phys. Rev. B {\bf 63}, 054514 (2001).
\bibitem{YOU} This is the case usually investigated (see
Refs.~\onlinecite{SHN}-\onlinecite{MAK}).
Self-inductance is taken into account in J.Q.~You,
C.-H.~Lam, and H.Z.~Zheng, Phys. ~Rev.~B {\bf 63}, 180501(R) (2001).
\bibitem{MEE} For an experimental multi-phonon generalization of the
Jaynes-Cummings model, see, {\it e.g.}, D.M. Meekhof, C. Monroe, B.E. King,
W.M. Itano, and D.J. Wineland, Phys. Rev. Lett. {\bf 76}, 1796 (1996).
\bibitem{teleportation} This is similar to the conditional quantum-phase gate
realized in a QED cavity for two photon qubits.\cite{TUR} Like the
controlled-NOT
gate, this conditional two-bit gate is also universal for
quantum computing.\cite{LLO,TUR} One can entangle charge qubits and
photons using this conditional quantum-phase gate (supplemented with
one-bit rotations) and employ photons as flying qubits to implement
quantum communications, {\it e.g.}, quantum teleportation.
\bibitem{rotation} One can irradiate the classical microwave field on a
charge qubit to realize the coupling between the charge qubit and the
classical field. Also, the classical microwave field can be adjusted to be
off-resonant with the QED cavity so as to eliminate its influence on the
cavity mode.
\bibitem{LLO} S.~Lloyd, Phys. ~Rev. ~Lett. ~{\bf 75}, 346
(1995); D.~Deutsch, A.~Barenco, and A.~Ekert, Proc. ~R. ~Soc. ~London,
Ser.~A {\bf 449}, 669 (1995).
\bibitem{array} A 2D array would require local gate
voltages $V_g(i,j)$ and local fluxes $\Phi_e(i,j)$ ({\it e.g.}, with
coils).
%
\end{references}
\end{document}